\documentclass[aps,prb,showpacs,floatfix,twocolumn,superscriptaddress]{revtex4-1}
\usepackage{graphicx}    
\usepackage{hyperref}    
\usepackage{bm}          
\usepackage[sumlimits,intlimits]{amsmath}
\usepackage{amsfonts,amssymb}
\usepackage{tikz}
\usepackage{cancel}
\usetikzlibrary{arrows, decorations.pathmorphing}
\pdfoutput=1

\begin{document}

\title{Plasmonic Hot Spots in Triangular Tapered Graphene Microcrystals}

\author{A. S. Rodin}
\affiliation{Boston University, 590 Commonwealth Ave., Boston MA 02215}
\author{Z. Fei}
\affiliation{University of California San Diego, 9500 Gilman Drive, La Jolla, California 92093}
\author{A. S. McLeod}
\affiliation{University of California San Diego, 9500 Gilman Drive, La Jolla, California 92093}
\author{M. Wagner}
\affiliation{University of California San Diego, 9500 Gilman Drive, La Jolla, California 92093}
\author{A. H. Castro Neto}
\affiliation{Boston University, 590 Commonwealth Ave., Boston MA 02215}
\affiliation{Graphene Research Centre and Department of Physics, National University of Singapore, 117542, Singapore}
\author{M. M. Fogler}
\affiliation{University of California San Diego, 9500 Gilman Drive, La Jolla, California 92093}
\author{D. N. Basov}
\affiliation{University of California San Diego, 9500 Gilman Drive, La Jolla, California 92093}

\date{\today}
\begin{abstract}
Recently, plasmons in graphene have been observed experimentally using scattering scanning near-field optical microscopy. In this paper, we develop a simplified analytical approach to describe the behavior in triangular samples. Replacing Coulomb interaction by a short-range one reduces the problem to a Helmholtz equation, amenable to analytical treatment. We demonstrate that even with our simplifications, the system still exhibits the key features seen in the experiment.
\end{abstract}

\pacs{
71.45.Gm 	
73.20.Mf 	
}

\maketitle

Recent scattering scanning near-field optical microscopy (s-SNOM) experiments on graphene in the mid-infrared spectral region have demonstrated the possibility of visualizing plasmonic oscillations in this material.~\cite{Chen2012oni, Fei2012gto} The behavior of plasmons is guided by the polarization function. From RPA formalism, it is predicted that the lifetime of plasmons is infinite.~\cite{Wunsch2006dpo} This is, however, contrary to the experimental observations which have demonstrated significant mode damping. The reason for this behavior can have different origins, such as residual conductivity,~\cite{Li2008dcd} arising from electron-electron interactions,~\cite{Mishchenko2008mci,Kotov2008eei,Koppens2011gpa,Kotov2012eei,Sodemann2012ict,Grigorenko2012gp} or disorder. This makes the problem significantly more complex. It has been shown~\cite{Fei2012gto} that that one can use a simplified hydrodynamic approach to circumvent these difficulties and qualitatively explain the main experimental findings. Within this formalism, the discreteness of electronic charge is traded for the continuous approximation so that one may use Ohm's law and the continuity equation to solve the problem. The two main geometries which appeared in the abovementioned literature were graphene half-planes and triangular microcrystals. Half-planes are samples that are much larger than the plasmon propagation length so that only one edge of the sample becomes relevant experimentally. From the theoretical standpoint, this simplifies the problem greatly as one can effectively reduce it to a single dimension. Our earlier work~\cite{Fei2012gto} focused on the formation of plasmonic effects in the half-plane geometry. While semi-infinite samples are much more common, we believe that the triangular microcrystals require further investigation due to the fascinating behavior that they manifest. In particular, a signature trait that is present in both Refs.~\onlinecite{Chen2012oni, Fei2012gto} is a series of unevenly-spaced bright spots along the edge of the microcrystal. These bright spots arise as a consequence of mode reflection and depend on the electronic properties of the sample. Even though triangular microcrystals are more rare than half-planes, they do have certain advantages. The half-plane theory~\cite{Fei2012gto} relies heavily on translational invariance of the system along the graphene edge: any inhomogeneities can act as additional reflectors, modifying the plasmonic behavior. In experiment, on the other hand, triangular systems are generally much smaller and this makes it easier to obtain a pristine sample. In this paper, our goal is to investigate plasmons in triangular geometries and explain the formation of signal maxima along the sample edges. Additionally, since this will turn out to be a geometric effect, our analysis also applies to other related phenomena, such as phonon-polaritons in similarly-shaped systems.

The system in question consists of an infinitely long triangular graphene microcrystal with the vertex angle $\theta_0$. The microcrystal is situated on a $\text{SiO}_2$ substrate and measurements are performed using SNOM in a manner described in Ref.~\onlinecite{Fei2012gto}. In short, an AFM tip is positioned above the microcrystal and polarized by a mid-infrared monochromatic laser at frequency $\omega$. This external potential interacts with the microcrystal, creating an induced potential. Finally, the scattered amplitude and phase are measured and recorded. The procedure is repeated for every point of the microcrystal, generating a signal map which captures the plasmonic behavior.

The total potential in the system consists of the external and the induced components. Assuming that the fields are too weak for non-linear effects, it is possible to write down the total potential in polar coordinates as the sum of the external and the induced ones:
\begin{equation}
\Phi(r,\theta) = \Phi_\text{ext}(r,\theta)+\Phi_\text{ind}(r,\theta)\,.
\label{eqn:tot_Phi}
\end{equation}
Utilizing Ohm's law, continuity equation, and Coulomb's law, one can express the induced potential in terms of the total one:
\begin{align}
-i\omega\rho+\nabla\cdot j &= 0\rightarrow \rho = \nabla\cdot\left[\frac{i\sigma}{\omega}\nabla\Phi(r,\theta)\right]\,,
\\
\Phi_\text{ind}(r,\theta)&=V\ast \nabla\cdot\left[\frac{i\sigma}{\omega}\nabla\Phi(r,\theta)\right]\,,
\label{eqn:Phi_ind}
\end{align}
where $\sigma$ is the momentum-independent conductivity and the convolution is taken with the interaction kernel $V$. Momentum independence of $\sigma$ is warranted by the fact that we are operating in the low-$q$ regime.~\cite{Wunsch2006dpo}  Recalling that the plasmon wave-vector at low $q$ is given by
\begin{equation}
q_p = i\kappa\omega/(2\pi\sigma)\,,
\end{equation}
the induced potential becomes
\begin{equation}
\Phi_\text{ind}(r,\theta)=-\frac{\kappa}{2\pi}V\ast \nabla\cdot\left[\frac{1}{q_p(r,\theta)}\nabla\Phi(r,\theta)\right]\,.
\label{eqn:tot_Phi_q}
\end{equation}
The term $\kappa$ is the mean dielectric constant of the media above and below graphene. In general, $\sigma$ is a function of frequency, but since we operate at a set laser wavelength, this does not present a problem. Aside from depending on $\omega$, $\sigma$ is also determined by the local charge density, which need not be constant. This means that the local plasmon wavelength can vary with position, a fact that has been demonstrated in our previous work.~\cite{Fei2012gto}

In an infinitely large system, setting the external potential to zero results in a plane-wave solution with the wave-vector $q_p$. For the system in question, however, the problem is more complicated and, instead of tackling it directly, we take a simplified approach. Experimentally, gaphene samples are not isolated systems. Other components, such as the substrate and the gate, can screen interactions in graphene. Assuming that the screening is strong enough, it is possible to replace the Coulomb term by a delta-function interaction which allows one to obtain an analytic solution to the problem. The benefit of using this substitution is the simplification of our integro-differential equation by turning it into a differential one. Taking the convolution in Eq.~\eqref{eqn:Phi_ind} and assuming a constant conductivity, one obtains:
\begin{equation}
\Phi_\text{ext}(r,\theta) = \Phi(r,\theta) + \frac{i\sigma}{\omega} \nabla^2\Phi(r,\theta)\,.
\label{eqn:Short_Plasmon}
\end{equation}
To make sure that the wavelength in this Helmholtz equation is the same as in the Coulomb case, we set $i\sigma/\omega = 1/q_p^2$. Since the Laplacian operator is separable in polar coordinates, we rewrite the external and total potentials as
\begin{align}
\Phi(r,\theta) &= \sum_n \phi_n(r)\cos\left(\nu_n\theta\right)\,,\quad\nu_n = n\frac{\pi}{\theta_0}\,,
\label{eqn:Phi_Harm}
\\
\Phi_\text{ext}(r,\theta) &= \sum_n \phi_n^\text{ext}(r)\cos\left(\nu_n\theta\right)\,.
\end{align}
The mode number $\nu_n$ ensures that the electric field normal to the edge vanishes to prevent the build up of charge. The next step is to rewrite Eq.~\eqref{eqn:Short_Plasmon} in the form of a Sturm-Liouville problem for individual harmonics:
\begin{align}
r q_p^2 \phi_n^\text{ext}(r) =&\frac{\partial}{\partial r}\left[r\frac{d\phi_n(r)}{dr}\right]+\frac{q_p^2r^2-\nu_n^2}{r}\phi_n(r)\,,
\label{eqn:Inhomo_Mode}
\\
\phi_n^\text{ext}(r)=&\frac{2-\delta_{n,0}}{\theta_0}\Re\left[\int_0^{\theta_0}d\theta'\,e^{i\nu_n\theta}\Phi_\text{ext}(r,\theta)\right]\,.
\label{eqn:Fourier_Ext}
\end{align}
The factor of $2$ appearing in Eq.~\eqref{eqn:Fourier_Ext} for $n\neq0$ terms arises because we have added positive and negative $n$ terms and from now on only consider $n\geq0$.

The solution to Eq.~\eqref{eqn:Inhomo_Mode} can be obtained using the method of Green's functions. Imposing the conditions that the derivative of the function vanishes at $r = 0$ and that for $r\rightarrow\infty$ the function behaves as a plane wave, one gets
\begin{align}
\mathcal{G}_n(r|\xi) &= -\frac{i\pi}{2}J_{\nu_n}\left(q_pr_<\right)H^{(1)}_{\nu_n}\left(q_pr_>\right)\,,
\\
 r_<& \equiv \min(r,\xi)\,,\quad r_> \equiv \max(r,\xi)\,,
\nonumber
\\
\phi_n(r)&=q_p^2\int_0^\infty d\xi\,\xi\,\mathcal{G}_n(r|\xi)\phi_n^\text{ext}(\xi)\,.
\label{eqn:Plasmon_Eqn_Gen}
\end{align}
In the expression above, $J_{\nu_n}$ is a Bessel function and $H^{(1)}_{\nu_n}$ is a Hankel function of the first kind.

\begin{figure}[h]
\includegraphics[width = 3.5in]{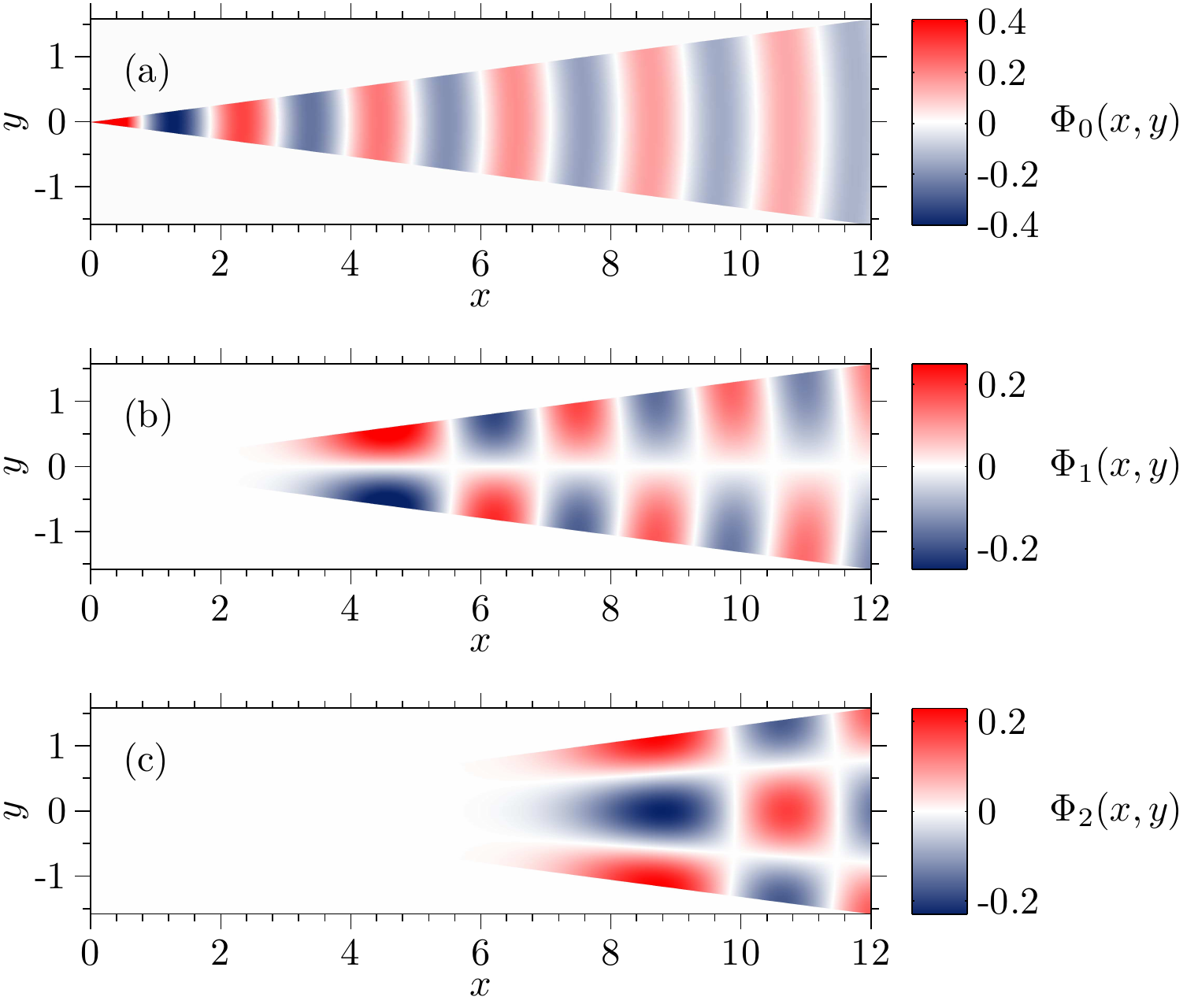}
\caption{First three modes ($n$ = 0, 1, 2) of the system for $q_p = 3$ and $\theta_0 = \pi/12$ given by Eq.~\eqref{eqn:modes}. These modes are stationary and their amplitude (illustrated by the colorbar) increases towads the apex of the microcrystal, governed by the Bessel function. For $n>1$, $J_n(q_p r)$ goes to zero as $r\rightarrow 0$, as is clearly seen from the vanishing amplitude for (b) and (c). Larger $n$ causes the amplitude to decay at higher $r$.}
\label{fig:Pure_Modes}
\end{figure}

During the course of the experiment, the polarized AFM tip couples to the modes of the system. Indeed, the essence of the Green's function approach is to calculate the responce of individual modes to the external perturbation. Therefore, to understand the observed behavior, one must gain an insight into the nature of these modes. From the solution of the differential equation, it is clear that they have the form of a Bessel function, multiplied by an azimuthal cosine term
\begin{equation}
\Phi_n(r,\theta) = J_n\left(q_pr\right)\cos\left(\nu_n\theta\right)\,,
\label{eqn:modes}
\end{equation}
see Fig.~\ref{fig:Pure_Modes}.

There are several important features immediately apparent from the illustration. First, the amplitude of the oscillations increases towards the apex of the microcrystal. The experimental consequence of this is that every mode has a stronger responce at low $r$. In addition, different modes can propagate to different $r$'s before they get reflected. This behavior is similar to that of waveguides where forward propagation depends on the perpendicular momentum. For this particular case, the total momentum of the plasmonic mode is the sum of radial and angular components:
\begin{equation}
q_p = \sqrt{k_\parallel^2 +\frac{\nu_n^2}{r^2}}\,.
\end{equation}
For $r < \nu_n/q_p$, $k_\parallel$ becomes imaginary and the mode is reflected. Interestingly, reflection points do not correspond to the bright spots seen in the experiment. Unlike the reflection points which are evenly spaced, according to Eq.~\eqref{eqn:Phi_Harm}, the distance between neighboring bright spots increases with larger $r$. This makes sense in the context of Bessel function modes as the coordinate of the first (and highest) maximum of $J_\nu(q_pr)$ Bessel function is approximately given by $q_pr\approx\nu+(\nu/2)^{1/3}$. This agrees with the experimentally observed spreading out of the signal maxima. In fact, one can move the reflection point for the modes by changing $q_p$ through gating or doping.

Having explored the underlying structure of the modes in the system, we return to the problem of determining the system's response to the external potential. Typically, the solution of Eq.~\eqref{eqn:Plasmon_Eqn_Gen} requires a numerical approach. Nevertheless, it is possible to get a feel for the structure of the solution by solving for the response to a delta-function perturbation $\delta(r-r_s)\delta(\theta-\theta_s)/r_s$:
\begin{equation}
\phi_n^\text{ext}(r) = \frac{2-\delta_{n,0}}{\theta_0} \frac{1}{r_s}\delta(r-r_s)\cos(\nu_n\theta_s)\,.
\end{equation}
Even though, in reality, SNOM picks up the electric field at at the location of the tip, here we focus on the potential under the tip. The reasoning is that the plasmonic features will be preserved in the $\Phi$ profile. Thus, we are interested in
\begin{equation}
\Phi(r_s,\theta_s) = \sum_n q_p^2\mathcal{G}_n(r_s|r_s) \frac{2-\delta_{n,0}}{\theta_0}\cos^2(\nu_n\theta_s)\,.
\end{equation}
For $n\rightarrow\infty$, $\mathcal{G}_n(r_s|r_s)\rightarrow-1/2\nu_n$. This leads to a divergence of the potential under the point tip. Therefore, we introduce a cutoff scale $\gamma = a / (a + r)$, where $a$ is the characteristic length of the external potential, and write:
\begin{equation}
\Phi(r_s,\theta_s) = \sum_n q_p^2\mathcal{G}_n(r_s|r_s)e^{-\gamma\nu_n} \frac{2-\delta_{n,0}}{\theta_0}\cos^2(\nu_n\theta_s)\,.
\label{eqn:mode_sum}
\end{equation}
Physically, this amounts to introducing a finite angular size to the potential.

\begin{figure}[h]
\includegraphics[width = 3.5in]{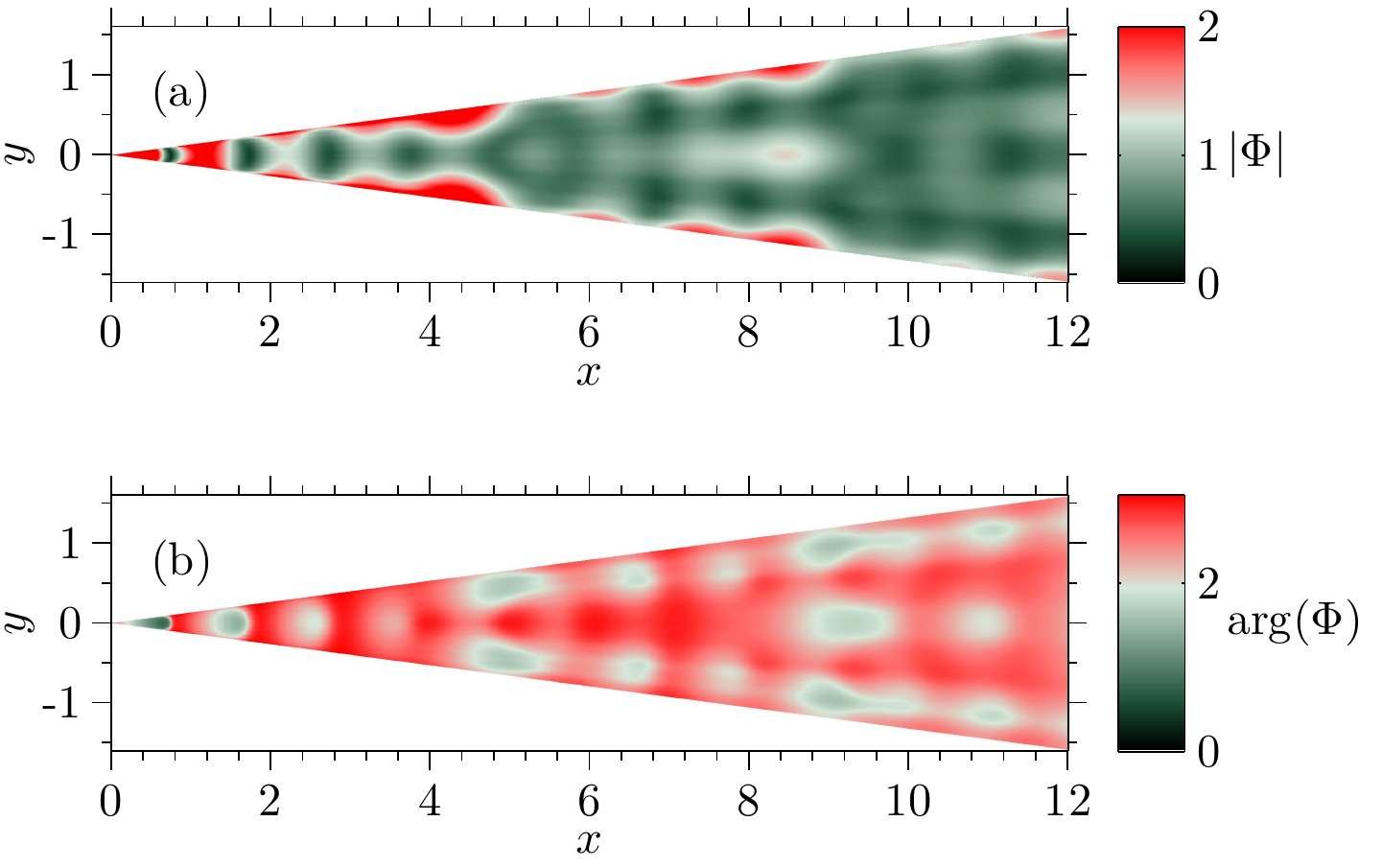}
\caption{Amplitude (a) and phase (b) of the total potential under the perturbing delta-function-like tip. The false color shows the amplitude in arbitrary units and the phase ranges between $0$ and $\pi$. Small damping was introduced to emulate the reduction of the amplitude away from the edges. The plot was obtained from Eq.~\eqref{eqn:mode_sum} for $\theta_0 = \pi/12$, $q_p = 3$, $\gamma = 1/20$. The summation is performed up to $n = 8$. The range of the response strength in (a) is cut off at 2 to remove the exceedingly high signal close to the apex and bring out the interference features.} 
\label{fig:flake_Delta}
\end{figure}

We have shown in Ref.~\onlinecite{Fei2012gto} that the plasmon momentum $q_p$ is a complex number due to the presence of damping in the system. While the amount of damping generally varies depending on the sample, in our earlier results we recorded $\Im[q_p]/\Re[q_p]\sim0.1$, which is the quantity that we adopt here. The result of the summation of the individual modes is given in Fig.~\ref{fig:flake_Delta}.

\begin{figure}[h]
\includegraphics[width = 3.5in]{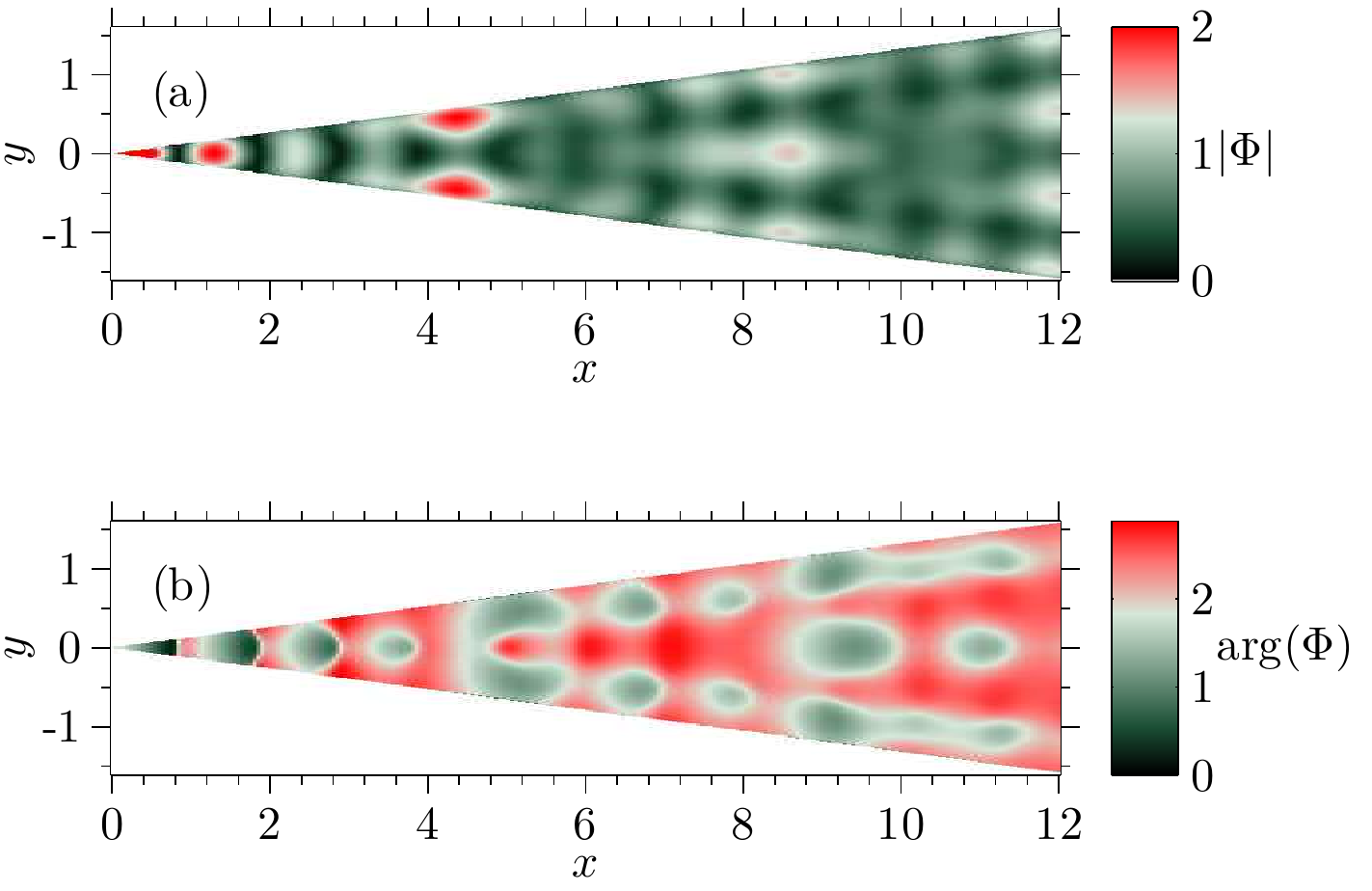}
\caption{Amplitude (a) and phase (b) of the total potential under the perturbing point-dipole potential of unit strength. The result is obtained using finite element method. The system is defined by $\theta_0 = \pi/12$, $q_p = 3$, and $\gamma = 1/20$. Small imaginary part is included in $q_p$ for damping. Just as in Fig.~\ref{fig:flake_Delta}, the false color shows (a) the signal strength in arbitrary units and (b) the phase ranging from 0 to $\pi$. In addition, the signal strength was renormalized and cut off at 2 in panel (a) to eliminate overly strong local response and to make the comparison with the delta-function perturbation easier.}
\label{fig:flake_Dipole}
\end{figure}

Despite the simplifications that we have utilized to obtain the result, one can see that the principal features, such as bright spots along the edge and alternating bright-dark lines parallel to it, are preserved. A closer approximation to the experimental result can be obtained by using a point-dipole perturbing potential to mimic the polarized tip, Fig.~\ref{fig:flake_Dipole}. Instead of employing the Green's function method, we use the standard finite element method to solve the problem. While the details of the result are different from the delta-function perturbation, the main features are found in both models. This means that our simplified approach captures the essence of the problem and provides the explanation for the bright spots.

We have demonstrated that it is possible to use simplified forms of charge interaction and SNOM potential to obtain a qualitatively correct picture of the plasmonic response in triangular microcrystals. At this point, we explain why our results are not surprising and, in fact, quite expected. Solving Eq.~\eqref{eqn:tot_Phi_q} for an infinitely large system in the absense of external potential results in plane-wave plasmon modes. Once the edges are included in the problem, the nature of the interaction between charges becomes important. This determines the finer details of the reflection, such as phase shift and a potential change in local wave vector. This would amount to solving the Helmholtz equation in Eq.~\eqref{eqn:Short_Plasmon} for a variable $q_p$ with an artificial phase change at the edge. This would still result in a solution composed of modes with different reflection points, giving rise to the bright spots along the edges. 

A.S.R. acknowledges DOE grant DE-FG02-08ER46512, ONR grant MURI N00014-09-1-1063. A.H.C.N. acknowledges NRF-CRP award "Novel 2D materials with tailored properties: beyond graphene" (R-144-000-295-281). Work at UCSD is supported by ONR.

%


\begin{thebibliography}{10}%
\makeatletter
\providecommand \@ifxundefined [1]{%
 \@ifx{#1\undefined}
}%
\providecommand \@ifnum [1]{%
 \ifnum #1\expandafter \@firstoftwo
 \else \expandafter \@secondoftwo
 \fi
}%
\providecommand \@ifx [1]{%
 \ifx #1\expandafter \@firstoftwo
 \else \expandafter \@secondoftwo
 \fi
}%
\providecommand \natexlab [1]{#1}%
\providecommand \enquote  [1]{``#1''}%
\providecommand \bibnamefont  [1]{#1}%
\providecommand \bibfnamefont [1]{#1}%
\providecommand \citenamefont [1]{#1}%
\providecommand \href@noop [0]{\@secondoftwo}%
\providecommand \href [0]{\begingroup \@sanitize@url \@href}%
\providecommand \@href[1]{\@@startlink{#1}\@@href}%
\providecommand \@@href[1]{\endgroup#1\@@endlink}%
\providecommand \@sanitize@url [0]{\catcode `\\12\catcode `\$12\catcode
  `\&12\catcode `\#12\catcode `\^12\catcode `\_12\catcode `\%12\relax}%
\providecommand \@@startlink[1]{}%
\providecommand \@@endlink[0]{}%
\providecommand \url  [0]{\begingroup\@sanitize@url \@url }%
\providecommand \@url [1]{\endgroup\@href {#1}{\urlprefix }}%
\providecommand \urlprefix  [0]{URL }%
\providecommand \Eprint [0]{\href }%
\providecommand \doibase [0]{http://dx.doi.org/}%
\providecommand \selectlanguage [0]{\@gobble}%
\providecommand \bibinfo  [0]{\@secondoftwo}%
\providecommand \bibfield  [0]{\@secondoftwo}%
\providecommand \translation [1]{[#1]}%
\providecommand \BibitemOpen [0]{}%
\providecommand \bibitemStop [0]{}%
\providecommand \bibitemNoStop [0]{.\EOS\space}%
\providecommand \EOS [0]{\spacefactor3000\relax}%
\providecommand \BibitemShut  [1]{\csname bibitem#1\endcsname}%
\let\auto@bib@innerbib\@empty
\bibitem [{\citenamefont {Chen}\ \emph {et~al.}(2012)\citenamefont {Chen},
  \citenamefont {Badioli}, \citenamefont {Alonso-Gonz\'{a}lez}, \citenamefont
  {Thongrattanasiri}, \citenamefont {Huth}, \citenamefont {Osmond},
  \citenamefont {Spasenovi\'{c}}, \citenamefont {Centeno}, \citenamefont
  {Pesquera}, \citenamefont {Godignon}, \citenamefont {Elorza}, \citenamefont
  {Camara}, \citenamefont {Garc\'{i}a~de Abajo}, \citenamefont {Hillenbrand},\
  and\ \citenamefont {Koppens}}]{Chen2012oni}%
  \BibitemOpen
  \bibfield  {author} {\bibinfo {author} {\bibfnamefont {J.}~\bibnamefont
  {Chen}}, \bibinfo {author} {\bibfnamefont {M.}~\bibnamefont {Badioli}},
  \bibinfo {author} {\bibfnamefont {P.}~\bibnamefont {Alonso-Gonz\'{a}lez}},
  \bibinfo {author} {\bibfnamefont {S.}~\bibnamefont {Thongrattanasiri}},
  \bibinfo {author} {\bibfnamefont {F.}~\bibnamefont {Huth}}, \bibinfo {author}
  {\bibfnamefont {J.}~\bibnamefont {Osmond}}, \bibinfo {author} {\bibfnamefont
  {M.}~\bibnamefont {Spasenovi\'{c}}}, \bibinfo {author} {\bibfnamefont
  {A.}~\bibnamefont {Centeno}}, \bibinfo {author} {\bibfnamefont
  {A.}~\bibnamefont {Pesquera}}, \bibinfo {author} {\bibfnamefont
  {P.}~\bibnamefont {Godignon}}, \bibinfo {author} {\bibfnamefont {A.~Z.}\
  \bibnamefont {Elorza}}, \bibinfo {author} {\bibfnamefont {N.}~\bibnamefont
  {Camara}}, \bibinfo {author} {\bibfnamefont {F.~J.}\ \bibnamefont
  {Garc\'{i}a~de Abajo}}, \bibinfo {author} {\bibfnamefont {R.}~\bibnamefont
  {Hillenbrand}}, \ and\ \bibinfo {author} {\bibfnamefont {F.~H.~L.}\
  \bibnamefont {Koppens}},\ }\href@noop {} {\bibfield  {journal} {\bibinfo
  {journal} {Nature}\ }\textbf {\bibinfo {volume} {487}},\ \bibinfo {pages}
  {77} (\bibinfo {year} {2012})}\BibitemShut {NoStop}%
\bibitem [{\citenamefont {Fei}\ \emph {et~al.}(2012)\citenamefont {Fei},
  \citenamefont {Rodin}, \citenamefont {Andreev}, \citenamefont {Bao},
  \citenamefont {McLeod}, \citenamefont {Wagner}, \citenamefont {Zhang},
  \citenamefont {Zhao}, \citenamefont {Thiemens}, \citenamefont {Dominguez},
  \citenamefont {Fogler}, \citenamefont {{Castro Neto}}, \citenamefont {Lau},
  \citenamefont {Keilmann},\ and\ \citenamefont {Basov}}]{Fei2012gto}%
  \BibitemOpen
  \bibfield  {author} {\bibinfo {author} {\bibfnamefont {Z.}~\bibnamefont
  {Fei}}, \bibinfo {author} {\bibfnamefont {A.~S.}\ \bibnamefont {Rodin}},
  \bibinfo {author} {\bibfnamefont {G.~O.}\ \bibnamefont {Andreev}}, \bibinfo
  {author} {\bibfnamefont {W.}~\bibnamefont {Bao}}, \bibinfo {author}
  {\bibfnamefont {A.~S.}\ \bibnamefont {McLeod}}, \bibinfo {author}
  {\bibfnamefont {M.}~\bibnamefont {Wagner}}, \bibinfo {author} {\bibfnamefont
  {L.~M.}\ \bibnamefont {Zhang}}, \bibinfo {author} {\bibfnamefont
  {Z.}~\bibnamefont {Zhao}}, \bibinfo {author} {\bibfnamefont {M.}~\bibnamefont
  {Thiemens}}, \bibinfo {author} {\bibfnamefont {G.}~\bibnamefont {Dominguez}},
  \bibinfo {author} {\bibfnamefont {M.~M.}\ \bibnamefont {Fogler}}, \bibinfo
  {author} {\bibfnamefont {A.}~\bibnamefont {{Castro Neto}}}, \bibinfo {author}
  {\bibfnamefont {C.~N.}\ \bibnamefont {Lau}}, \bibinfo {author} {\bibfnamefont
  {F.}~\bibnamefont {Keilmann}}, \ and\ \bibinfo {author} {\bibfnamefont
  {D.~N.}\ \bibnamefont {Basov}},\ }\href@noop {} {\bibfield  {journal}
  {\bibinfo  {journal} {Nature}\ }\textbf {\bibinfo {volume} {487}},\ \bibinfo
  {pages} {82} (\bibinfo {year} {2012})}\BibitemShut {NoStop}%
\bibitem [{\citenamefont {Wunsch}\ \emph {et~al.}(2006)\citenamefont {Wunsch},
  \citenamefont {Stauber}, \citenamefont {Sols},\ and\ \citenamefont
  {Guinea}}]{Wunsch2006dpo}%
  \BibitemOpen
  \bibfield  {author} {\bibinfo {author} {\bibfnamefont {B.}~\bibnamefont
  {Wunsch}}, \bibinfo {author} {\bibfnamefont {T.}~\bibnamefont {Stauber}},
  \bibinfo {author} {\bibfnamefont {F.}~\bibnamefont {Sols}}, \ and\ \bibinfo
  {author} {\bibfnamefont {F.}~\bibnamefont {Guinea}},\ }\href@noop {}
  {\bibfield  {journal} {\bibinfo  {journal} {New J. Phys.}\ }\textbf {\bibinfo
  {volume} {8}},\ \bibinfo {pages} {318} (\bibinfo {year} {2006})}\BibitemShut
  {NoStop}%
\bibitem [{\citenamefont {Li}\ \emph {et~al.}(2008)\citenamefont {Li},
  \citenamefont {Henriksen}, \citenamefont {Jiang}, \citenamefont {Hao},
  \citenamefont {Martin}, \citenamefont {Kim}, \citenamefont {Stormer},\ and\
  \citenamefont {Basov}}]{Li2008dcd}%
  \BibitemOpen
  \bibfield  {author} {\bibinfo {author} {\bibfnamefont {Z.~Q.}\ \bibnamefont
  {Li}}, \bibinfo {author} {\bibfnamefont {E.~A.}\ \bibnamefont {Henriksen}},
  \bibinfo {author} {\bibfnamefont {Z.}~\bibnamefont {Jiang}}, \bibinfo
  {author} {\bibfnamefont {Z.}~\bibnamefont {Hao}}, \bibinfo {author}
  {\bibfnamefont {M.~C.}\ \bibnamefont {Martin}}, \bibinfo {author}
  {\bibfnamefont {P.}~\bibnamefont {Kim}}, \bibinfo {author} {\bibfnamefont
  {H.~L.}\ \bibnamefont {Stormer}}, \ and\ \bibinfo {author} {\bibfnamefont
  {D.~N.}\ \bibnamefont {Basov}},\ }\href@noop {} {\bibfield  {journal}
  {\bibinfo  {journal} {Nature Physics}\ }\textbf {\bibinfo {volume} {4}},\
  \bibinfo {pages} {532} (\bibinfo {year} {2008})}\BibitemShut {NoStop}%
\bibitem [{\citenamefont {Mishchenko}(2008)}]{Mishchenko2008mci}%
  \BibitemOpen
  \bibfield  {author} {\bibinfo {author} {\bibfnamefont {E.~G.}\ \bibnamefont
  {Mishchenko}},\ }\href@noop {} {\bibfield  {journal} {\bibinfo  {journal}
  {Europhys. Lett.}\ }\textbf {\bibinfo {volume} {83}},\ \bibinfo {pages}
  {17005} (\bibinfo {year} {2008})}\BibitemShut {NoStop}%
\bibitem [{\citenamefont {Kotov}\ \emph {et~al.}(2008)\citenamefont {Kotov},
  \citenamefont {Uchoa},\ and\ \citenamefont {{Castro Neto}}}]{Kotov2008eei}%
  \BibitemOpen
  \bibfield  {author} {\bibinfo {author} {\bibfnamefont {V.~N.}\ \bibnamefont
  {Kotov}}, \bibinfo {author} {\bibfnamefont {B.}~\bibnamefont {Uchoa}}, \ and\
  \bibinfo {author} {\bibfnamefont {A.~H.}\ \bibnamefont {{Castro Neto}}},\
  }\href@noop {} {\bibfield  {journal} {\bibinfo  {journal} {Phys. Rev. B}\
  }\textbf {\bibinfo {volume} {78}},\ \bibinfo {pages} {035119} (\bibinfo
  {year} {2008})}\BibitemShut {NoStop}%
\bibitem [{\citenamefont {Koppens}\ \emph {et~al.}(2011)\citenamefont
  {Koppens}, \citenamefont {Chang},\ and\ \citenamefont {Garcia~de
  Abajo}}]{Koppens2011gpa}%
  \BibitemOpen
  \bibfield  {author} {\bibinfo {author} {\bibfnamefont {F.~H.~L.}\
  \bibnamefont {Koppens}}, \bibinfo {author} {\bibfnamefont {D.~E.}\
  \bibnamefont {Chang}}, \ and\ \bibinfo {author} {\bibfnamefont {F.~J.}\
  \bibnamefont {Garcia~de Abajo}},\ }\href@noop {} {\bibfield  {journal}
  {\bibinfo  {journal} {Nano Lett.}\ }\textbf {\bibinfo {volume} {11}},\
  \bibinfo {pages} {3370} (\bibinfo {year} {2011})}\BibitemShut {NoStop}%
\bibitem [{\citenamefont {Kotov}\ \emph {et~al.}(2012)\citenamefont {Kotov},
  \citenamefont {Uchoa}, \citenamefont {Pereira}, \citenamefont {Guinea},\ and\
  \citenamefont {{Castro Neto}}}]{Kotov2012eei}%
  \BibitemOpen
  \bibfield  {author} {\bibinfo {author} {\bibfnamefont {V.~N.}\ \bibnamefont
  {Kotov}}, \bibinfo {author} {\bibfnamefont {B.}~\bibnamefont {Uchoa}},
  \bibinfo {author} {\bibfnamefont {V.~M.}\ \bibnamefont {Pereira}}, \bibinfo
  {author} {\bibfnamefont {F.}~\bibnamefont {Guinea}}, \ and\ \bibinfo {author}
  {\bibfnamefont {A.~H.}\ \bibnamefont {{Castro Neto}}},\ }\href@noop {}
  {\bibfield  {journal} {\bibinfo  {journal} {Rev. Mod. Phys.}\ }\textbf
  {\bibinfo {volume} {84}},\ \bibinfo {pages} {1067} (\bibinfo {year}
  {2012})}\BibitemShut {NoStop}%
\bibitem [{\citenamefont {Sodemann}\ and\ \citenamefont
  {Fogler}(2012)}]{Sodemann2012ict}%
  \BibitemOpen
  \bibfield  {author} {\bibinfo {author} {\bibfnamefont {I.}~\bibnamefont
  {Sodemann}}\ and\ \bibinfo {author} {\bibfnamefont {M.~M.}\ \bibnamefont
  {Fogler}},\ }\href@noop {} {\bibfield  {journal} {\bibinfo  {journal} {Phys.
  Rev. B}\ }\textbf {\bibinfo {volume} {86}},\ \bibinfo {pages} {115408}
  (\bibinfo {year} {2012})}\BibitemShut {NoStop}%
\bibitem [{\citenamefont {Grigorenko}\ \emph {et~al.}(2012)\citenamefont
  {Grigorenko}, \citenamefont {Polini},\ and\ \citenamefont
  {Novoselov}}]{Grigorenko2012gp}%
  \BibitemOpen
  \bibfield  {author} {\bibinfo {author} {\bibfnamefont {A.~N.}\ \bibnamefont
  {Grigorenko}}, \bibinfo {author} {\bibfnamefont {M.}~\bibnamefont {Polini}},
  \ and\ \bibinfo {author} {\bibfnamefont {K.~S.}\ \bibnamefont {Novoselov}},\
  }\href@noop {} {\bibfield  {journal} {\bibinfo  {journal} {Nat. Photonics}\
  }\textbf {\bibinfo {volume} {6}},\ \bibinfo {pages} {749} (\bibinfo {year}
  {2012})}\BibitemShut {NoStop}%
\end{thebibliography}
\end{document}